\documentclass[sigconf,nonacm]{acmart}

\AtBeginDocument{%
  \providecommand\BibTeX{{%
    \normalfont B\kern-0.5em{\scshape i\kern-0.25em b}\kern-0.8em\TeX}}}

\setcopyright{rightsretained}
\acmConference[CSCW Workshop '20]{CSCW '20: Reconsidering Scale and Scaling in CSCW Research}{October 17--18, 2020}{Virtual}
\begin{document}

\title[Against Scale]{Against Scale: Provocations and Resistances to Scale Thinking}

\author{Alex Hanna}
\authornotemark[1]
\email{alexhanna@google.com}
\affiliation{Google Research}
\orcid{0000-0002-8957-0813}
\author{Tina M. Park}
\authornote{Both authors contributed equally to this research.}
\email{tina.m.park@gmail.com}
\affiliation{Independent Researcher}

%\renewcommand{\shortauthors}{Hanna and Park}

% <concept_id>10003456</concept_id>
% <concept_desc>Social and professional topics</concept_desc>

% \ccsdesc[500]{Social and professional topics}

% \keywords{scalability, CSCW}

\maketitle

\section{Introduction}

At the heart of what drives the bulk of innovation and activity in Silicon Valley and elsewhere is {\it scalability}. Scalability is that most-important, desirable attribute that “connotes the ability of a system to accommodate an increasing number of elements or objects, to process growing volumes of work gracefully, and/or be susceptible to enlargement” \cite]p.~195]{Bondi2000}. It means that a small start-up is capable of growing into a multinational corporation, serving a global audience in a short period of time. This unwavering commitment to scalability - to identify strategies for efficient growth - is at the heart of what we refer to as “scale thinking.” Whether people are aware of it or not, scale thinking is all-encompassing. It is not just an attribute of one’s product, service, or company, but frames how one thinks about the world (what constitutes it and how it can be observed and measured), its problems (what is a problem worth solving versus not), and the possible technological fixes for those problems. 

We argue that scale thinking represents a very particular form of growth, a form that has become the default and predominant approach within the current technology sector. Scale thinking presumes that everything can be made more efficient -- that products and services can be supplied and consumed at faster and faster speeds, for more and more people. Such efficiency is possible when production of the good or service can be expanded without significantly modifying the inputs (e.g., the type of raw material used, the type of labor required) or infrastructure necessary. Such growth is indisputably a good thing. For developers and investors, it means capturing greater market shares with lower levels of investment. Even those committed to social good over financial gain tend to believe that the efficient use of resources and the rapid expansion of potential solutions would help deal with the world’s ills faster and more effectively -- they are able to do more good with fewer resources.

This paper examines different facets of scale thinking and its implication on how we view technology and collaborative work. We argue that technological solutions grounded in scale thinking are unlikely to be as liberatory or effective at deep, systemic change as their purveyors imagine. As we will discuss, the logics that drive scale thinking are antithetical to constructing solutions which can produce systemic, equity-driven social change. Rather, solutions which {\it resist} scale thinking are necessary to undo the social structures which lie at the heart of social inequality. We draw on recent work on mutual aid networks and propose questions to ask of collaborative work systems as a means to evaluate technological solutions and identify resistances to scale thinking.

\section{Prior Literature on Scale \& What is Scale Thinking?}

“Scale thinking” is an approach that centers on and prioritizes {\it scalability}\footnote{Our intervention borrows heavily from Lilly Irani's \cite{irani2018} critique of "design thinking," which characterizes the concept as a type of racialized Western expertise that resists the imaginary mechanistic reasoning of a looming Asian labor threat. It also borrows from Ruha Benjamin's critique of design as a "colonizing project" which "submerg[es] so much heterogeneity" under its name \cite[pp. 175-76]{benjamin2019}.}. Scalability refers to the ability of a system to expand without having to change itself in substantive ways or rethinking its constitutive elements \cite{Tsing2012}. Scalability goes hand-in-hand with other macro-level processes of making modernity legible, including standardization \cite{timmermans2010}, classification \cite{bowkerStar2000}, and colonial trade \cite{Tsing2012}. Investors seek technological start-ups which are designed to grow quickly \---\ businesses that are structured to meet any level of demand at moment’s notice.  In an essay titled "Startup = Growth," Y Combinator founder Paul Graham discusses how the distinguishing feature of startups is its ability to grow, stating "not every newly founded company is a startup" and most of those are in service. The startup is an instantiation of scale thinking which requires immediate growth. "A barbershop doesn't scale," he quips. In scale thinking, this feature of any given system is assumed to be the most relevant for success and longevity. As Werner Vogels, chief technology officer of Amazon, wrote, “scalability cannot be an after-thought. It requires applications and platforms to be designed with scaling in mind…” \cite{Vogels2006}. 

By centering scalability in the design and development process, scale thinking revolves around three key tenets: (1) Scalability is a morally good quality of a system; (2) Quantification is a necessary part of designing scalable systems; and (3) Scalability is achieved by identifying and manipulating quantifiable, core elements or attributes of a system. 

First, in scale thinking, scalability is taken from being a desirable trait to a {\it morally good} trait. Insofar as capitalism values the maximization of production and the minimization of waste -- whether it is unused labor or raw materials and goods -- scalability is imbued with a moral goodness because it centers on designing a system that is able to serve a greater number of people with fewer resources over time. When a system is scalable, growth -- which itself is fundamentally a positive and good thing -- can take place without the loss of time, resources, or effort. The absence of scalability is equated with wastefulness: opportunities to meet the needs of an ever-growing body of people are lost and precious resources (money, time, raw materials) are utilized ineffectively. Especially at a time when natural resources are dwindling, while populations are growing, wastefulness is marked as evil and morally corrupt. It becomes taken-for-granted that to be a “good” company or person, one should {\it obviously} seek scalability. 

Moreover, scalability is a prerequisite to technical implementation; accordingly, solutions which don't scale are morally abject. Large tech firms spend much of their time hiring developers who can envision solutions which can be implemented algorithmically. Code and algorithms which scale poorly are seen as undesirable and inefficient. Many of the most groundbreaking infrastructural developments in big tech have been those which increase scalability, such as Google File System (and subsequently the MapReduce computing schema) and distributed and federated machine learning models. Large-scale technical systems operate much like the markets that preceded them, and are considered by liberals to be in and of themselves a moralizing and virtuous force \cite{FourcadeHealy}.

Given the importance of growth to scalability, the measurement of growth and efficiency is paramount to measuring the worth and value of the scalable system. The greater the growth, the more efficient the use of inputs and resources, and the greater the capture of demand, the better and more valuable the system. In scale thinking, developing tactics for measuring such growth is as important as designing for scalability itself (because if no one sees the growth, did it really happen?). Quantification, or the production and act of transforming human observations and experiences into quantities based on a common metric \cite{EspelandStevens} is an important procedural aspect of scalability. 

Scalability is achieved when a system is able to expand {\it without rethinking basic elements} \cite{Tsing2012}. The system is designed in such a way that can accommodate new inputs without changing its fundamental framework. To create such a system, thus, requires an understanding of what those core elements or attributes of the system are -- of what aspects can remain unchanged and what elements can be changed to accommodate a growing system. Improving efficiency also means eliminating excess and removing non-essential elements to ensure the system runs smoothly. To engage in scale thinking, then, is to try and reduce a complex process or interaction to its most elemental and simplistic exchange of input to output.

\section{Implications of scale thinking}

The implications of scale thinking are three fold. In the first, scale thinking requires units of work to be interchangeable, abstract, and universal. Units themselves can be servers, offices, databases, individual workers. Next, scale thinking requires users to be of the same kind, or within a narrow set of predefined bounds, which is most often to the detriment of people "at the margins," marked by systems of racism, transphobia, and ableism. Lastly, scale thinking has implications for the legibility of users, with a natural endpoint in data gathering and datafication of individuals.

Scale thinking demands a standardization of inputs and outputs. Global supply chains demand modularity: shipping containers require only a pickup point and destination \cite{Posner2018}. As {\it The Atlantic} contributor Alexis Madrigal has suggested, containers are the "embodiment...of global capitalism." \cite{Madrigal2017}. What's significant is how readily that embodiment has translated to technological work.

The container metaphor readily extends to the realm of software development and deployment. Docker, the first widely used "container" service, allows developers to develop "standard unit[s] of software that package up code" and run them in different environments. Docker's logo is a whale with a set of shipping containers set on top of it. Kubernetes, likewise, is a software package used to manage "deployment, scaling, and management of containerized applications." Accordingly, its logo is a ship's wheel, steering containers to where they need to go.

\begin{figure}
    \includegraphics[scale=0.5]{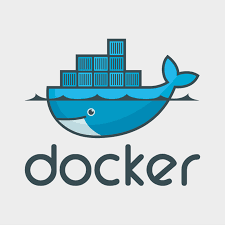}
    \includegraphics[scale=0.5]{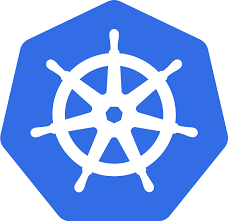}
    \caption{The Docker and Kubernetes logos, respectively.}
\end{figure}

The ubiquity of the container metaphor highlights the embeddedness of scale thinking in modern computing and infrastructural development; the metaphor extends to how developers think about work units, software teams, and technological organizations. If modern software companies are supposed to be "flat," then the work teams within them operate as standardized containers that take as input design and requirement documents, and output code, processes, and product. These organizations scale with the addition of more containerized work teams. At the base unit of this operation, the individual tech worker needs to have such standardized inputs and outputs. As such, scaling has a fractal quality within tech development.

While scale thinking permeates the organization of the production side of tech development, it sharply determines its consumptive side as well. Scale thinking requires a sameness of user, or a user which falls within a tightly bound constraint of imagination. In a containerized world, interchangeability is critical to the operation working, and users operate in a standardized manner. The desire of the startup is to ensure that users fall within the bounds of the universal. Heterogeneity becomes antithetical to scalability, because the same product/service can no longer be duplicated to sufficiently serve a suffuse audience. A varied user base means that many different solutions are needed, rather than a scalable solution. Despite Graham's cry to "do things that don't scale," \cite{Graham2012} a startup's outputs need to be constrained insofar as they do just that.

The failures for users that fall outside of universality of scaled solutions abound. Media scholar Safiya Noble accounts in stark detail the racist and sexist failures of web search for queries for Black, Latina, and Asian women and girls \cite{Noble2018}. In the book's conclusion, she details how Kandis, the owner of a Black-serving salon in a college town of a predominantly white institution, had to contend with Yelp and the erasure of her business: "I quickly realized that Internet/Yelp told people that I did not exist" \cite]p.~175]{Noble2018}. Kandis goes on to detail the extensive hoops she has to go through to get listed on the service, and the hesitance of her primarily Black clientele to "check-in" using Yelp's functionality, because of their sense of already being over surveilled. 

Datafication, then, becomes the endpoint of scale thinking. Much has been made of the impulse of datafication of the individual \cite{VanDjick, CheneyLippold} and its anti-Black, carceral dimensions \cite{Browne, benjamin2019}. Datafication's move -- and therefore scale thinking's move -- is to find ways to rationalize the individual into legible data points. A single data point is rarely useful on its own -- data points only matter insofar as they accumulate and move through the world as a new form of capital \cite{Sadowski}. This can only be made possible via creation of systems undergirded by scale thinking, and the building of systems which operate as massive accrual machines.

\section{Resistances to Scale Thinking}
In April 2020, at the height of the Coronavirus pandemic in New York City, George Farcasiu, the "lead automation engineer" of the mutual aid network Bed-Stuy Strong told VICE's technology-focused publication {\it Motherboard} that “We’re making sure we’re building tools that are about organizing people to interact with neighbors, not treating volunteers as boxes that do work.” He continued, “We’re not looking for a system that scales. We’re looking at the social system we have and augmenting and enabling that” \cite{Rose}.

On the flip side of this, tech companies continue to struggle with diversity and inclusion among their employees, despite years of investment and highly publicized efforts. One early strategy adopted by many tech companies was the installation of “people analytics” teams with human resources divisions to provide “data-driven,” scalable solutions \cite{Bersin}. Efforts focused on scalable tactics -- focusing on hiring percentages and inclusion metrics, implementing bias workshops, developing employee resource groups -- with little to no impact on the composition of the workforce \cite{RooneyKhorram} or the experiences of marginalized people within the company \cite{Tiku}. 

These two examples illustrate two opposite, but active resistances to scale thinking. In the former, the resistance is centered on the hopeful and local in mutual aid networks. In the latter, however, we find resistance anchored in the pessimistic and anti-social. To conclude this piece, we draw on recent writings on mutual aid to discuss the ways in which scale limits -- and actively inhibits -- participation in tech and society. Indeed, scale thinking forces particular types of participation to operate as extractive or exploitative labor \cite{Sloane2020}. We ask readers to consider what potential resistances to scale thinking may look like, and invite them to think through what kinds of technology encourage collaborative work in ways which don't replicate its logics. 

One such framework is that of mutual aid. Mutual aid allows a possible way to think of collaborative work more fruitfully. Critical legal scholar Dean Spade defines mutual aid as: 

\begin{quote}
a form of political participation in which people take responsibility for caring for one another and changing political conditions, not just through symbolic acts or putting pressure on their representatives in government but by actually building new social relations that are more survivable. \cite[p.~136]{Spade}
\end{quote}

The concept of mutual aid has been around for years (many trace the concept to the anarchist theorist Peter Kropotkin) and the practice for much longer. It is a particular tradition of relationship building and has many local instantiations, including Black mutual aid traditions in the diaspora and the Black Panthers breakfast program in Oakland to name a few \cite{Harris, Heynen, Mochama}. Unlike other social assistance programs, mutual aid is oriented around collective problem-solving to develop strategies and obtain resources to help a community of people meet one another’s needs while simultaneously organizing towards different social and political relations, both among community members and between the community and other structures of power. It is the latter that distinguishes mutual aid from forms of charity or assistance: assistance provided between group members is not only about sharing the resources on hand, but changing what resources are made available and how they are distributed in general; it is about shifting dynamics of power. 

For example, for the Black Panther Party, the Breakfast Program was an important mutual aid program that was integral to their overall political strategy to transform U.S. politics. In addition to providing necessary support and aid for the communities they served, it also functioned as a model that the federal government could use to establish a national school breakfast program \cite{Heynen}. The urgency of the COVID pandemic has revived interest and networks around mutual aid. In the US, with the lack of federal and state intervention resulting in mass infections and death, mutual aid groups have emerged as a means to provide material relief for vulnerable populations, including Black and brown, elderly, low-income, disabled, and queer and transgender people.

Mutual aid networks, by their nature, are not intended to “scale”. While scale thinking emphasizes abstraction and modularity, mutual aid networks encourage concretization and connection. Mutual aid is intended to operate as a mode of radical collective care \cite[p.~131]{Spade} in which individuals in the network have their direct material needs met, regardless of considerations of those receiving aid falling into a set of datatified categories, such as "deserving or undeserving." While scale thinking encourages top-down coordination, mutual aid considers building skills for "collaboration, participation, and decision making" \cite[p.~137]{Spade}. Mutual aid focuses on "how to organize human activity without coercion" \cite{Spade}. And mutual aid encourages building of solidarities between people with different needs. While mutual aid is not the only framework through which we can consider a move away from scale thinking-based collaborative work arrangements, we find it to be a fruitful one to theorize and pursue.

The Bed-Stuy neighborhood in New York City’s Brooklyn borough reflects gentrification trends. It is a mix of newly arrived residents who are white, more financially-secure, technologically-oriented, both professionally and personally, and long-time residents who are Black or Latinx, older, and more likely to be negatively harmed by COVID-19 and its related economic impacts \cite{NYSComptroller}.\footnote{It is also a neighborhood with a rich history of localized Black activism that opposed its designation as the “largest ghetto” \cite{Woodsworth}.} And, given the disparate impacts of COVID-19 on Black and Brown communities and lower-income communities, the desire to make the aid distribution system nimble enough to grow quickly and across more neighborhoods, if not entire cities and regions, is a well-intentioned one. Like many other “tech for social good” projects, it is not unreasonable to fixate on the end goal of the project - in this case, distributing money and food to as many people as possible - in the hopes of assisting the most number of people. 

Yet, the developers who volunteered to build out the tech-enabled aspect of the work resisted building a system focused on scalability, explicitly focusing on developing a system that centers both the recipients of the aid and the volunteers who distribute the aid. Instead, they looked to using technology to augment the relationships formed between volunteers and recipients. As noted by Alyssa Dizon, a civic technologist and a volunteer with Bed-Stuy Strong:
\begin{quote}
The most important piece of our operation is our relationship to the neighborhood, our sensitivity to the community, our approach, our values, our culture as an organization. The tech is a layer of it, and it’s not the most important layer of it at all.” \cite{McKenzie}
\end{quote}

The priority for developing the technological infrastructure for Bed-Stuy Strong is focused less on building a system that can expand quickly to serve the most people possible. Instead, it is focused on easing and strengthening social connections between the people involved in doing the mutual aid: the volunteers, the donors, and the recipients. The technology layer is built with the understanding that those involved, rather than being easily replaceable units, are not homogenous. Organizers presume that the people and their relationships with one another will change, including their roles within the mutual aid network. A volunteer may be a donor one week, and due to changes in circumstances, may become a recipient the next as well. People who receive aid may also volunteer their time to provide aid to others.

Furthermore, there is an expectation that the users of the mutual aid system are different and have different needs. Rather than building for a homogenous base of users or clients, the tech-oriented developers within Bed-Stuy Strong oriented themselves towards identifying technology that could help them identify differences between users, such as those who are elderly, immuno-compromised, living with children, or in need of immediate assistance. By doing so, they can render assistance in a way that is most functional and appropriate for the recipient. Rather than trying to identify users to who fit, or could be made to fit, into the system, they adapted the work of volunteers and donors to meet the needs of a vast array of recipients.

We conclude this section with a few considerations around how to organize collaborative work in technological systems. Spade proposes four questions in assessing reforms and tactics which we use as our jump-off point: 

\begin{quote}
Does [the reform or tactic] provide material relief? Does it leave out an especially marginalized part of the affected group (e.g., people with criminal records, people without immigration status)? Does it legitimize or expand a system we are trying to dismantle? Does it mobilize people, especially those most directly impacted, for ongoing struggle? \cite[p.~133]{Spade}
\end{quote}

We add three more, related questions for technical collaborative work systems: Does the technological system centralize power (either through coordination, data extraction, or authority) or distribute it between developers and users? Does the technological system treat the contributions and experiences of individuals as interchangeable or as uniquely essential? Lastly, does it open up avenues for participation, and are those avenues of participation mobilizing or demobilizing?

\section{Conclusion}

In this piece, we identify "scale thinking" as a mode of thinking prevalent in Silicon Valley and other technological development centers. We discuss the dimensions and implications of scale thinking, and how this type of thinking is universally valorized within technological development. We then introduce resistances to scale thinking, namely mutual aid, as an alternative mode of arranging social relations, and offer several provocations and questions for designers and theorists of cooperative and collaborative work systems. 

Pandemics and disasters are sites of immense loss of life, property, and livelihood. They also starkly highlight what elements, at their root, are not working for most people and contribute to interlocking systems of inequality. Failures under disasters are often failures of technological scale and the attendant scale thinking which undergirds them. We hope that this set of provocations introduces a mode of evaluating whether systems created by designers and other tech workers are working for those who need them most.

\bibliographystyle{ACM-Reference-Format}
\bibliography{base}

%\appendix

\end{document}